\documentclass[12p,preprint,amsmath,showpacs,superscriptaddress,nofootinbib]{revtex4-1}
\usepackage{graphicx,epstopdf,epsfig,psfrag}

\begin{document}

\title{Possibility to create a coronavirus sensor using an optically excited electrical signal}

\author{Ognyan Ivanov}
\email{ogi124@yahoo.com, tel: +359 89 99 77 880  }
\author{Peter Todorov}
\author{Zhivko Stoyanov}
\affiliation{Georgi Nadjakov Institute of Solid State Physics, Bulgarian Academy of Sciences, 
	72~Tzarigradsco~Chaussee~Blvd., Sofia, 1784, Bulgaria 
}

\begin{abstract}
	The research of our team has led to a series of new experimental results, revealing the so-called electromagnetic echo effect (EMEE). The EMEE signal is highly sensitive even to small changes in the composition and properties of the object under study~-~gas, liquid, or solid. The method is fast and contactless, provides real-time analysis, and can be considered universal.
	We have an idea to create a sensor to detect COVID-19 in three cases - in the air, on hard surfaces, and in fluids from the human body. The approach involves detecting specific reactions to the virus, even if they are invisible without specialized equipment. Our experience shows that the EMEE signal is very sensitive to weak or practically imperceptible reactions. Therefore, we believe that it is quite possible to register the presence of COVID-19.
	We have many experimental results in this regard that support this idea. We have experience in creating sensors to monitor the processes, composition, and properties of fluids (including biological ones), as well as to detect pollution and contamination in the atmosphere.
\end{abstract}


\maketitle

\section*{Introduction}

In recent years, researchers from the Institute of Solid State Physics (ISSP), Bulgarian Academy of Sciences, discovered and developed the so-called Electromagnetic Echo Effect (EMEE)~\cite{o6}, previously known as surface photo charge effect~\cite{o1,o4}.
The effect is based on the interaction of any solid body with an electromagnetic field, which generates in the body an alternating, electric signal with the same frequency as the frequency of the incident field~\cite{o11}.

As EMEE voltage can be generated for every type of matter, the method offers convenient, fast and accurate measurements, and might be applied to virtually any substance. Last but not least, EMEE-based devices are rather cost-effective, as compared with the conventional ones, in terms of both investment and running costs, being thereby capable of low-cost sensing tests.

The need for finding suitable sensors for detection of coronaviruses has become more pressing, as the number of patients with severe respiratory syndromes has increased over the last 10 years, with viruses such as SARS-CoV, MERS-CoV and, more recently, SARS-CoV2 or better known as COVID-19~\cite{b31} responsible for most of them.

The ISSP research team elaborates a work plan to create and implement sensors for coronaviruses.

\section*{Sensors based on electromagnetic echo effect}

According to the experimental results, EMEE is induced by electromagnetic field irradiation with frequency in not only the visible and the adjacent regions of the spectrum, but also with frequencies ranging from 1 Hz up to 1 GHz. The effect is expected to occur in the entire electromagnetic spectral range. No studies have been done in the area between 1 GHz and infrared and for frequencies higher than the ultraviolet range. In these cases, expensive specialised equipment is needed. At low electromagnetic field frequencies, the measurements can be performed at the emission frequency. In the visible and adjacent spectral range, an additional amplitude modulation of the applied radiation is used. It is needed because it is still not possible to perform direct measurement of signals with frequencies in the higher GHz and THz ranges.

When a constant electric field is applied, EMEE is not observed. This peculiarity can be used for a simple and prompt verification whether the measured signal is due to EMEE or to other similar effects, such as external and internal photoelectric effects, thermal electricity, etc. There are other ways to prove the case of EMEE. Signals generated by EMEE or by the external photoelectric effect can be distinguished by using, for example, spectral studies.

The differentiation is possible because a limit exists for the wavelength of the incident radiation in the case of the external photoelectric effect, after which it cannot occur. Beyond this limit, only EMEE can be observed~~\cite{o7}.

Physical models regarding the mechanism of EMEE in various types of solids were created in order to explain the experimental results. For example, in conductors, the effect manifests as the occurrence of voltage in the solid sample under irradiation, which is presumed to be caused by the redistribution of near-surface charges induced by the electromagnetic field.  The voltage originates from changes induced in the electrostatic potential of the double layer at the conductor surface.  Overall, currently there is still no comprehensive theoretical model describing the effect in all solids – dielectrics, semiconductors, and metals~\cite{o4,o8}.

An important property that has been established is that each solid generates a specific signal, which is determined by inherent properties of the object. For example, generally speaking, each body can be characterized by its weight, determined by the interaction between the body and the gravitational field of the Earth. In a similar way, it can be characterized by EMEE, which is determined by the interaction of the body with an electromagnetic field.

\begin{figure}
	\centering
	\includegraphics[width=.95\linewidth]{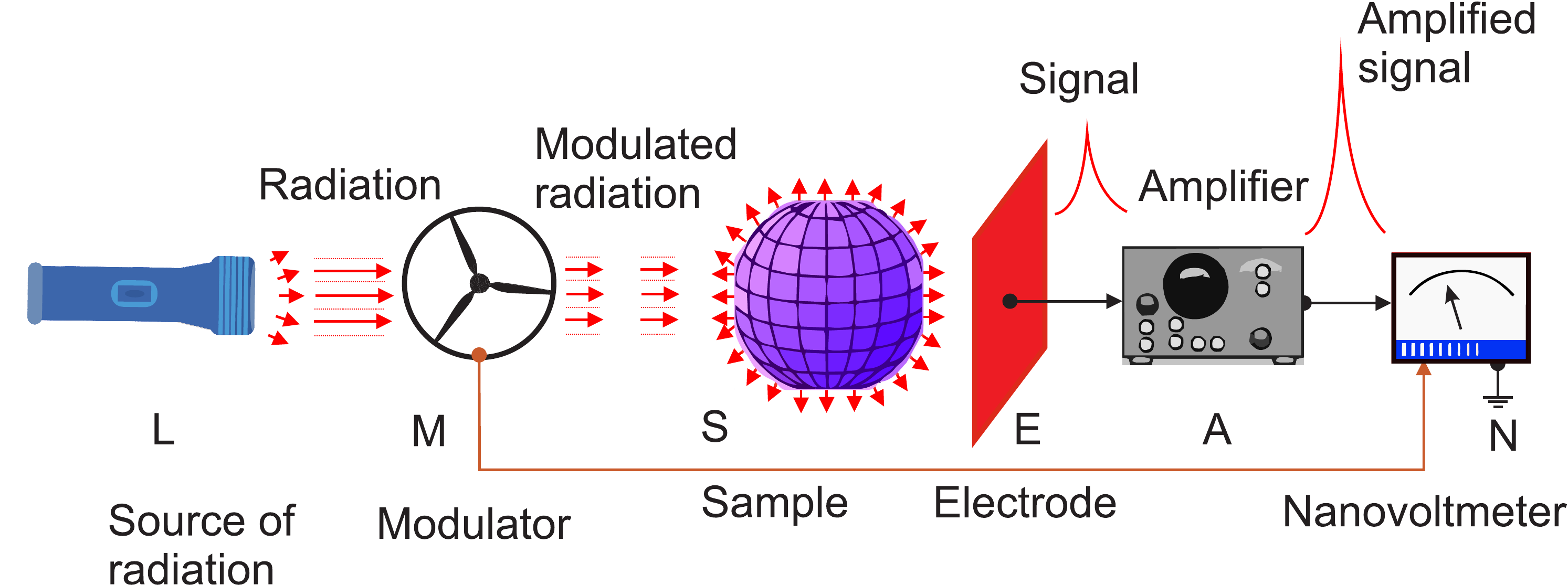}
	\caption{Experimental setup for EMEE observation: L - light source; M - modulator; S - measuring structure; 
		E – electrode; A - amplifier; N - lock-in nanovoltmeter.
	}
	\label{fig:EMEESET}
\end{figure}

An experimental setup for EMEE investigation in the visible range of the spectrum is shown in figure \ref{fig:EMEESET}.
The source (L) of incident radiation generates in a wide spectral range (white light) or a monochromatic one (laser). The most convenient way is to work with a laser. The incident light beam is chopped into periodic pulses using a modulator (M). A pulsed laser, a pulsed LED or another modulated light source could also be used instead of such a modulator. The studied sample (S) is placed in the measuring arrangement. The latter is specifically developed for the conducted research. It must ensure reliable fastening of the sample providing maximum strength and protection to the signal against interfering electromagnetic fields. The preamplifier (A) amplifies the obtained signal. The detected signal has a very low amplitude, thus a special nanovoltmeter (N), for example a lock-in nanovoltmeter, capable of extracting the signal from the background noise, is used. The modulator (M) supplies the reference signal to the nanovoltmeter. With the help of the reference signal, by using a phase-sensitive detection, interferences can be eliminated so that only the useful signal is measured.

Numerous practical applications of EMEE have been experimentally proven. Some of them are:
contactless characterization of semiconductors~\cite{o8};
monitoring processes taking place in fluids~\cite{o15, o11, o15, o14, o17} including biological ones~\cite{o19};
retrieving information about an irradiated surface in terms of defects, irregularities and impurities~\cite{o11};
monitoring of gasoline octane-number~\cite{o17};
express contactless determination of the chemical composition of test samples (for example, false coins, various types of absorbing filters for gases and liquids can be monitored to determine when they need to be replaced, testing of drinking water from the public water distribution system~\cite{o11}, etc.);
control of phase transitions in liquid crystals~\cite{o30};
monitoring the quality of raw materials~\cite{o11};
control of fog parameters, including the presence of impurities in them~\cite{o26}; etc.

\section*{Possibility to create sensors detecting coronaviruses based on the EMEE}

\begin{figure}[h]
	\centering
	\includegraphics[width=\linewidth]{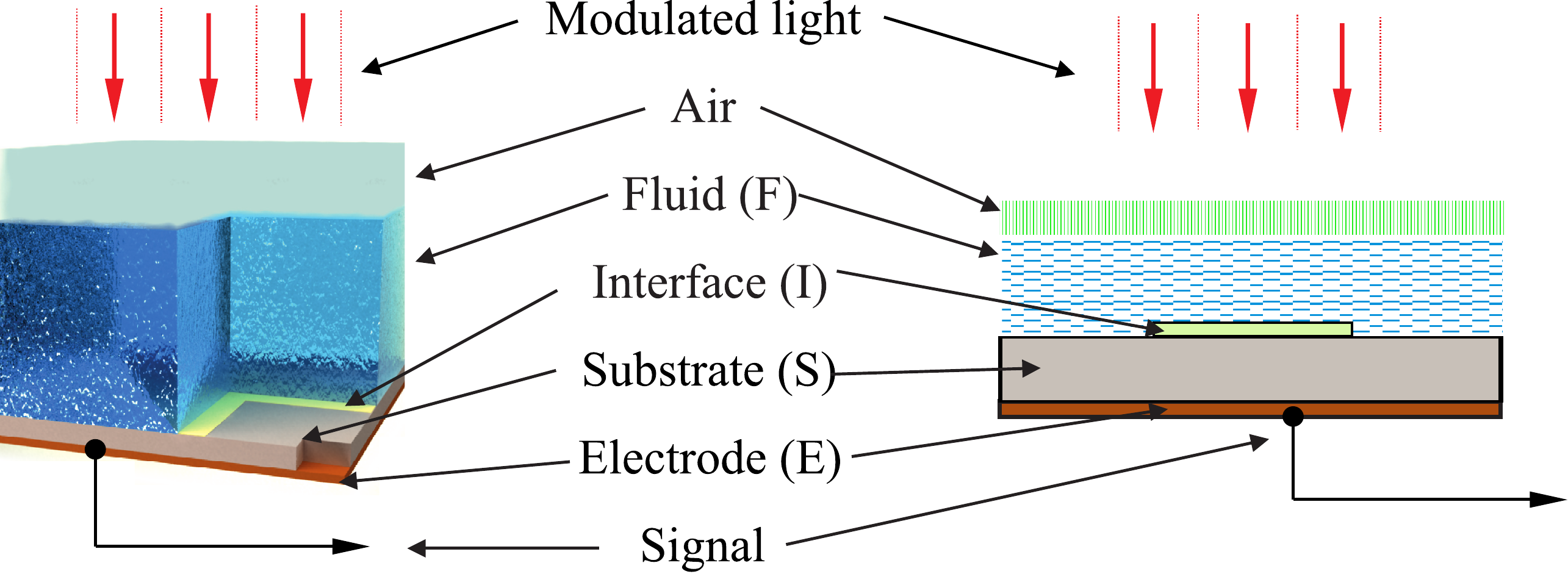}
	\caption{Possible arrangement of a EMEE - based sensor for fluids: S - solid; I - solid liquid interface, generating the signal; F - fluid under study; E – electrode.
	}
	\label{fig:Fluid}
\end{figure}

During our research, it became apparent that gas, liquid or vapour sensors, based on the electromagnetic echo effect, present a very attractive possibility for practical applications. The concept of the set-up is the same for all types of fluids: the surface of the irradiated solid should be in contact with the fluid under investigation. Moreover, between them there will be a boundary surface – an interface (the boundary between the solid and the fluid). Any variations in the fluid characteristics will induce a corresponding change in the boundary surface~– especially in the area irradiated by the incident radiation.
The measured signal is generated in this exact area. As a result, a change in the EMEE signal corresponding to changes in the fluid can be registered. 
It has been proven that even small changes in the controlled fluid can induce measurable variations in the EMEE signal, since the electron properties of the surface are substantially influenced by the solid-fluid interface. Therefore, changes in fluid properties can be detected, provided that all other conditions remain constant. 
The idea is illustrated in figure \ref{fig:Fluid}, where (S) is the irradiated solid (generally, it should be a substrate generating a strong EMEE signal, and this signal should be strongly dependent on changes in the fluid under study), (F) is the studied fluid that forms an interface (I) with the solid in the spot of irradiation. 

The EMEE signal is measured by an electrode (E) and using the appropriate apparatus. In the case of gas or vapour sensors, it is advisable to use as a substrate a solid that has maximal adsorption capacity to the respective fluid.  

Provided that a suitable measuring structure is selected, minimal amounts of liquid (even a small drop), are sufficient for the analysis. Such a structure is described in~\cite{o17}. In this case, the interface that generates the signal is formed by that part of the drop that interacts with the solid.

The experiments carried out with various liquids, gases and vapours have provided proof for the feasibility of the above considerations. 
The experimental data has shown that the EMEE signal is highly sensitive to variations in the composition and the properties of the investigated fluid~\cite{o15, o11, o19, o17, o30, o26, o14}. 
The proposed method can be used to study various fluids under different conditions and can be considered universal. The measurements are rapid and contactless, and offer real-time analysis. For example, in liquids, a characteristic feature of EMEE is that the AC signal, due to the effect, has a unique value for each liquid. Any minor change in the liquid (concentration, contamination, pre-treatment, etc.) influences the EMEE signal, indicating that the processes taking place at the interface depend strongly on the liquid characteristics~\cite{o15}.

Together with the already existing methods for control, EMEE can also be prospectively used for a complementary, rapid analysis. The main advantage of such an analytical method is its universal nature – since the EMEE voltage can be generated for every type of fluid, the method can be applied for characterization of any fluid. The combination of optical probing of the sample and electrical detection of the generated signal is also of great importance, as it provides convenient, fast and precise measurements. In addition, the implementation of the method can be easily realised from technical point of view and does not involve significant financial investments.

The work in creating a sensor for coronaviruses could proceed in three directions at the same time: detection of the virus in the air, on surfaces and in a fluid from the human body.
A possible approach is the following: finding a surface (liquid or solid) that is sensitive to the emergence of the virus, i.e. when the virus comes into contact with the surface, a reaction will occur, even if this reaction is imperceptible without specialized equipment. Our experience has shown that the EMEE signal is very sensitive to weak or practically imperceptible reactions occurring at the interface. Therefore, we believe that it is completely feasible to register the presence of the controlled component (in this case - COVID-19).

While we are presenting some examples of possible reactions, our investigations will not be limited only to them.
Studied samples (e.g. human body liquids) might be treated by a reagent to catalyse a specific virus-sensitive reaction. Thus, the first objective is to detect animal viruses of the coronavirus type. Furthermore, COVID-19 will specifically be addressed by developing a reliable coronavirus detection biosensor whose response is strictly specific to this virus type. For that purpose, the research team intends to firstly produce a sensor for detecting well-studied antigen-antibody responses. Antibodies that specifically interact with the coronavirus spike protein are thus vital~\cite{b34}. 

Unlike immunologically mediated methods, the anticipated biosensor will not necessarily chase antibodies in humans, but will control for the presence of viral particles in the airways secretions. Coronavirus antibodies are commercially available today and can also be insulated from the blood plasma or serum of individuals recovered from the SARS-CoV-2 infection. Alternatively, monoclonal antibodies can be employed, which may be a safe way of highly specific detection of various coronavirus family members~\cite{b31}. It is also possible to try detecting the agglutination reaction, which accompanies the antigen-antibody reaction, as an aggregation of insoluble antigen and IgM and/or IgG antibodies, subject to availability of electrolytes and a suitable temperature. When the antigen and antibody concentrations are equalised, visible aggregates are formed, which makes it much easier to detect such agglutination complexes~\cite{b38}. The thermodynamic changes during agglutination can also be efficiently detected to expose the existence of coronaviruses. Other possible reagents, capable of stimulating a response, will actively be sought as well.

The new experimental system that will be developed will generally include: an optical part, a source of the studied samples and reagents, a sensory structure and a registration block. In the sensor structure, a contact between the irradiated solid surface and the investigated samples will be implemented. This structure will generate the measured signals. It will be possible to attach to it additional elements such as temperature scanning system, backlight with different wavelengths, source of additional electrical signals, etc. In addition, an important element of the experimental system will be the registering block that needs to be adequately equipped with instruments for measuring and recording alternating electrical signals with amplitudes of the order of microvolts and nanovolts. These signals must be measured even in the presence of a significantly strong electrical interference. The most likely approach is to use phase sensitive technique. Reference signal for it will be provided by the optical block. It may be necessary for the optical block to provide scanning across the wavelength range.

\section*{Conclusion}

The development of a sensor based on the electromagnetic echo effect, for fast detection of viruses (COVID-19 or others depending on the settings) without the need for expensive consumables will strongly support any actions against diseases that constitute public health risk. The new technique is expected to provide a completely new and robust way to prevent strong negative influences on society when viral outbreaks occur.

The measured EMEE signal for fluids is formed by the fluid – solid interface. The essence of the idea is to put the irradiated solid surface in contact with the fluid in which a virus might be present. A reagent that can generate a response specific to the controlled virus type can also be added. Since the electron properties of the solid surface are influenced by the adjacent fluid layer, one can expect that optically excited changes in such a system will induce measurable EMEE signals generated due to the presence of a specific virus. When all other conditions are constant, changes caused by the presence of this virus can be detected. The presence of the virus will significantly affect the properties of the illuminated solid – fluid interface and, thus, the signal from the sensory structure. Therefore, EMEE can be used for development of new techniques for virus presence control. The approach is valid for detecting viruses in the human body, in the air, and on solid surfaces. 

The idea outlined above presents great possibilities because it combines optical probing of the sample with electrical detection of the generated signal. Some of the advantages will be instantaneous results and little to no expensive consumables. Such devices will be cheap to produce and small enough to be used in field work. The possible use of a laser (if necessary to work with such a radiation source) as a light source will not increase the size and weight of the device, since currently there are very small and compact laser modules with internal intensity modulation.

\section*{Acknowledgements}

This work has been funded by FP7-SEC-2012-1 program of the EU Commission under grant number 312804 and partially supported by the Bulgarian Ministry of Education and Science under the National Research Programme "Young scientists and postdoctoral students" approved by DCM~$\sharp$~577~/~17.08.2018.
The authors will like to thank Kostadin Fikiin, Stoyan Shishkov, Venelin Tsvetkov and Betina Hristova for the fruitful discussions.

\bibliography{EMEE-Arxiv_2020}

\end{document}